\documentclass[runningheads]{cl2emult}
\usepackage{amsmath}
\usepackage{psfig}
\newcommand{\avida}{{\sf avida\ }}
\newcommand{\D}{{\boldsymbol D}}
\begin{document}

\title*{Evolution of Genetic Organization \\ in Digital Organisms}

\author{Charles Ofria \and Christoph Adami}
\institute{Beckman Institute and Kellogg Radiation Laboratory
\\California Institute of
  Technology\\Pasadena, CA 91125}

\maketitle

\begin{abstract}
  We examine the evolution of expression patterns and the organization
  of genetic information in populations of self-replicating digital
  organisms.  Seeding the experiments with a linearly expressed
  ancestor, we witness the development of complex, parallel secondary
  expression patterns.  Using principles from information
  theory, we demonstrate an evolutionary pressure towards overlapping
  expressions causing variation (and hence further evolution) to
  sharply drop.  Finally, we compare the overlapping sections of dominant
  genomes to those portions which are singly expressed and observe a 
  significant difference in the entropy of their encoding.
\end{abstract}

\section{Introduction}
Life on Earth is the product of approximately four billion years of
evolution, with the vast majority of beginning and intermediate states
lost to us forever.  The exact details of how we evolved to become what
we are may be impossible to ascertain for sure, but we may still be
able to better
understand the evolutionary pressures exerted on life, and from that
reconstruct sections of the path our evolution is likely to have taken.

Here we look at a fundamental issue to life as we know it; the
organization of the genetic code and the differentiation in its
expression.  DNA is structured into many distinct genes which can be
concurrently active, transcribed and expressed in an asynchronous,
(i.e., differentiated) manner.  Extant living systems have evolved to
a state where multiple genes influence each other, typically without
sharing genetic material.  It appears that in all higher life forms
each gene has its own unique position on the genome, while the
transcription products often interact with unique positions
``downstream''.  Those organisms which do exhibit {\em overlapping}
expression patterns are mostly virii and bacteriophages~\cite{NORM83}.
This suggests that genomes containing only purely localized,
non-overlapping genes must have evolved later on~\cite{KEESE92}.
    
Upon initial inspection, the reason for a spatially separated layout
appears uncertain.  A modular design may be quite common in
artificially created coding schemes such as computer programs, but, in fact,
only reflects a designer's quest to create human-understandable
structures.  Evolution has no such incentive, and will always exert
pressure towards the most immediate solution given the current
circumstances.  A more compressed coding scheme, perhaps with
overlapping genes, would allow a sufficiently shorter code
that would minimize the mutational load and hence be able to preserve
its information with a higher degree of accuracy. Furthermore,
such overlapping regions might be used for gene regulation.  Why this is
not much more common becomes clearer when we observe those examples
from nature where these overlapping reading frames do exist, such as
DNA phages~\cite{NORM83} and eukaryotic viruses~\cite{SAM89}.
Even in these organisms only some sections of code overlap, but
examination of those sections reveals that they contain little
variation---almost all of the nucleotides are effectively frozen in
their current state from one generation to the
next~\cite{MIYA78,MIZO97}.  This occurs because for any mutation to be
neutral in such a section of genetic code, it must be neutral to {\it
  both} of the genes which it would affect.  Further, most of the
mutations that occur in DNA which are neutral occur in the third
nucleotide of a codon, as substitutions in that position are often synonymous.
When overlapping genes have offset (out-of-phase) reading frames, however,
the position of the third nucleotide in one gene maps to the first or
second in the other, leaving no redundancy.

We have investigated the development of genome organization and
differentiation in {\em digital organisms}: populations of
self-replicating computer-code living in a computer's 
memory. Such ``Artificial Life'' systems have proven to be useful
test cases to investigate the biochemical paradigm because the
computational chemistry the digital organisms are based on share {\em
  Turing universality} with their biochemical cousins, i.e., just as
any type of organism appears to be implementable in biochemistry, the
digital organisms can in principle compute any (partially-recursive)
function~\cite{IAL98}. Due to the ease with which experiments can be prepared, 
data can be gathered, and trials can be repeated, digital organisms
present an important tool to study universal traits in the evolution
and development of symbolic sequences. Differentiation in digital
organisms was first investigated within the {\sf tierra}
architecture~\cite{RAY91,RAY94,RAY98} and we comment on those results below.

For the present study, we have extended our \avida
system~\cite{IAL98} to allow for the expression of a {\em second} gene
to occur in parallel.  We
then processed the evolution of 600 populations from a seed program to
complex information-processing sequences for an average of over 9000
generations each.  The 600 trials were divided into four sets which
differ in the length of the seed program, constraints on size
evolution, and their ability to express multiple portions of code in
parallel.  All populations with a genetic basis allowing for the
development of multiple {\em threads} learn to use them almost
immediately (each thread is an instruction pointer which executes the
code independently), but the methods by which this happens are quite
distinct and varied. In the next section, we outline the most
important design characteristics of the \avida system, focusing mostly
on the particular experimental setup needed for this study. Also, we
outline the kind of observables which we record, and discuss measures
of differentiation. In Section 3 we present results obtained with our
multiple-expression digital chemistry and compare them to controls in
which no secondary expression was allowed.  In Section 4 we study the
evolution of differentiation for different experimental boundary
conditions, while Section 5 explores in more detail the organization
and development of genes at the hand of an example.  We close in
Section 6 with a discussion of the evidence and conclusions, and issue
caveats about applying the lessons learned directly to biochemistry.

\section{Experimental Details}

\subsection{The {\sf Avida} Platform}
The computer program \avida is an auto-adaptive genetic system~\cite{ADA95} 
designed primarily for use as a platform in Artificial Life research.
The system consists of a population of self-reproducing strings of
instructions with a Turing-complete genetic basis subjected to
Poisson-random mutations during reproduction.  The population adapts
to a combination of an intrinsic fitness criterion
(self-reproduction) and an externally imposed (extrinsic) fitness
landscape provided by the researcher by creating an information-rich
environment. 

A normal \avida organism is a computer program written in a very
simple assembly language, with 28 possible commands for each line
(Table I).
\begin{table}[h]
\centering
\caption{Standard (single expression) \avida instruction set\label{mnem}}
\vskip 0.5cm
\begin{tabular}{|c|c|}
\hline Instruction type& Mnemonic\\ \hline
flow control &
{\tt jump-b}, {\tt jump-f}, {\tt call}, {\tt return}\\

conditionals &
{\tt if-n-eq}, {\tt if-less}, {\tt if-bit-1}\\

self analysis &
{\tt search-f}, {\tt search-b}\\

computation &{\tt shift-l}, {\tt shift-r}, {\tt inc}, {\tt dec}, {\tt
  swap}\\
 &  {\tt swap-stk}, {\tt push}, {\tt pop}, {\tt add}, 
  {\tt sub}, {\tt nand}\\
metabolic &{\tt alloc}, {\tt divide}, {\tt copy}\\
I/O & {\tt get}, {\tt put} \\
labels & {\tt
  nop-A}, {\tt nop-B}, {\tt nop-C}\\ \hline
\end{tabular}
\end{table}

These programs exist on a two-dimensional lattice with toroidal
boundary conditions, and are executed on simple virtual CPUs residing
at the lattice-sites which process their code
allowing them to interact with their environment and perform functions
such as self-replication, as well computations on numbers which are
found in the external environment.  For more details on the virtual
CPUs in {\sf avida}, see~\cite{OBA98}.

In order to study the evolution of code expression, we have extended
the instruction set of Table I to allow for more than one instruction
pointer to execute a program's code.  Within the biochemical
metaphor, the {\em simultaneous} execution of code is viewed as the
concurrent expression of two genes, i.e., the chemical action of two
proteins. The first new instruction allows a program to initiate a new
expression: {\tt fork-th}. Its execution creates a new instruction
pointer (``forking off a thread'') which immediately executes the next
instruction, while the original thread skips it.  Thus, {\tt fork-th}
is the rough equivalent of a promoter sequence in biochemistry.  In a
sense, this secondary expression is rather trivial and leads to
redundancy; if the second thread is not sufficiently altered by the
instruction following the {\tt fork-th}, it simply executes the
identical code as the first thread in lock-step.  Of course, we are
interested in how the organisms use this redundancy as a starting
point to {\em diversify} the expression.

The second new instruction {\em inhibits} an expression:
{\tt kill-th} removes the instruction pointer which executed it, while
the third addition {\tt id-th} {\em identifies} which pointer is currently
executing the code, i.e., which pattern is currently being expressed.
We expect the three commands together to be useful in the {\em
  regulation} of expression. In principle, more than two instruction
pointers can be generated by repeated issuing of the {\tt fork-th}
command, but here we restrict ourselves to a maximum of two threads in
order not to complicate the analysis. In nature, of course, complex
genomes express hundreds of proteins simultaneously.

As our experiments begin with a self-replicating program which does
not use any of the multiple expression commands, the first question
might be whether or not multiple expression will develop at all. In
fact, it does almost instantly, as secondary expression (typically in
the trivial mode mentioned earlier) appears to be immediately
beneficial, perhaps in the same manner as simple gene doubling or a
second promoter sequence.  From here on, differentiation evolves,
i.e., the two instruction pointers begin to adapt independently, to
express more and more different code.  Ultimately one might expect
that each pointer executes an entirely different section of code,
achieving local separation of genes and fully parallelized execution.
The mode and manner in which this
separation occurs is the subject of this investigation.

Several hundred independent experimental trials and controls were
obtained in this study, testing different experimental conditions.
For each of these trials we keep a record of a variety of statistics,
including the dominant genotype at each time step, from which we can
track the progression of evolution of the population, in particular by
 studying the details of its expression patterns.

\subsection{Basic Analysis Metrics}\mbox{}\\

In order to track the differentiation of the threads, we need to
develop a means to monitor the divergence between the two instruction
pointers roaming the genome. Also, to study the evolutionary pressures
such as the {\it mutational load}, we need to introduce some standard
(and some less standard) observables which allow us to track the {\em
  adaptability} of the population. This is one of the major advantages
of digital chemistries---some of the data that we collect is
impossible to accurately obtain in biochemical systems, and even less
practical to analyze.

{\bf Fitness} is measured as the number of offspring a genome produces
per unit time, normalized to the replication rate of the ancestor.
Thus, in all experiments the fitness of the dominant genotype starts
at one and increases.  Fitness improvements are due to two effects:
the optimization of the gene for replication (the ``copy-loop'')
leading to a smaller gestation time, as well as the development of new
genes which accomplish computations on externally provided random
numbers.  These computations are viewed as the equivalent of
exothermic catalytic reactions mediated by the expression products. We
reward the accomplishment of all bit-wise logical operations performed
on up to three numbers by speeding-up the successful organism's CPU at
a rate commensurate to the difficulty of the computation.

{\bf Fidelity} is the probability for an organism to produce an
offspring perfectly identical to itself, i.e., the probability that
the offspring is unaffected by mutations during the copy process.
For pure copy-mutations (each instruction copied is mutated with a
probability $R_c$)
\begin{equation}
    F = (1 - R_c)^\ell
\end{equation}
where $\ell$ is the organism's sequence length.
In an adapting population, other factors can affect
the fidelity and lead to low-fidelity organisms even while the
theoretical fidelity is high. On the other hand, the development of
error-correction schemes could increase the actual fidelity.
    
{\bf Neutrality} $\nu$ is the probability that an organism's fitness
is unaffected by a single point mutation in its genome.  This is
calculated by obtaining all possible one-point mutations of the
examined genome, and processing each of them in isolation to determine
fitness. The neutrality is then the number of neutral mutations
divided by the total tested:
\begin{equation}
\nu=\frac{N_{\rm neut}}{\ell(\D-1)}\;,
\end{equation}
where $\D$ is the number of different instructions in the digital
chemistry, i.e., the size of the instruction set. 

The preceding three indicators are key in determining the ability of
an organism to thrive in an \avida environment. Fitness, fidelity, and
neutrality correspond respectively to an organism's ability to create
offspring, for those offspring to have a minimum mutational load, and
for them to survive those mutations which they do bear.  Apart from
this, however, there is another aspect which is necessary for a
phylogenetic branch to be successful, and that is its ability to
further adapt to its environment.  To characterize this, we define two
more genomic attributes:

{\bf Neutral Fidelity} is a measure which can be calculated once an
organism's neutrality is known.  It is the probability that an
organism will give birth to an identical {\em or equivalent}
offspring.  Taking $f_c= R_c (1 - \nu)$ to be the probability for a line
to be mutated {\em and} be non-neutral to the organism, we obtain the
neutral fidelity as:
\begin{equation}
    F_{\rm neut} = (1 - f_c)^\ell\;.
\end{equation}

{\bf Genomic Diffusion Rate} is the probability for an offspring to
have a genome {\em different} from its parent, but to be otherwise
equivalent (i.e., neutral.) This is obtained by subtracting
the genome's fidelity from its neutral fidelity
\begin{equation}
D_g = F_{\rm neut} - F\;.
\end{equation}
This is a particularly important indicator as it is the rate at which
new, viable genotypes are being created, which in turn is the pace at
which genetic space is being explored, and therefore directly
proportional to the rate of adaptation.

\subsection{Differentiation Measures}\mbox{}\\

The following measures and indicators keep track of code-differentiation. In 
biochemistry, the differentiation of
expression can be very varied, and includes overlapping reading frames
(in-phase and out-of phase), overlapping operons and promoter
sequences, and gene regulation.
Obviously, there are no reading frames in our digital chemistry, but
it is possible for a sequence of instructions to give rise to a
different computation depending on which thread is executing it, in
particular if one gene contains another (as is very common in
overlapping biochemical genes~\cite{MBOG}).  Also,
thread-identification may lead one thread to execute instructions
which are skipped by the other thread, and threads may interact to
turn each other on and off---a case of digital gene regulation. All
such differentiation however has to evolve from the trivial secondary
expression discussed earlier, and we consequently need to
monitor the divergence of thread-execution with suitable measures.

{\bf Expression Distance} is a metric we use to determine the
divergence of the two instruction pointers.
Simply put, this measurement is the average {\em distance} (in units
of instructions) between the sections of the genome actively being expressed
by the individual threads.  At the initial point leading to secondary
expression, this distance is zero as the two threads execute the same
code in lock-step.  If this value is high relative to the length of
the genome, it is a strong indication that the instruction pointers
are expressing different sections of the genetic code at any one time,
while if it is low, they most likely move together with identical (or
nearly so) execution patterns.  However, this measure only indicates
the differentiation between execution at a particular point in {\em
  time}, implying that if the execution is simply time-offset, this
metric may be misleading.

{\bf Expression Differentiation} distinguishes execution patterns with
characteristically differing {\em behavior}.  Each execution thread is
recorded with time, and a count is kept of how {\em often} each
portion of the genome is expressed.  The expression differentiation is
the fraction of the genome in which those counts differ. Thus, the
ordering of execution (time-delay) is irrelevant for this metric; only
whether the code ends up getting expressed {\em differently} by one
thread vs. the other is important. 

\subsection{Information Theoretic Measures}\mbox{}\\

We use information theory in order to distinguish sequences which do
or do not code for genes. In our digital chemistry, regions which do
not code for a gene are either {\em unexecuted}, i.e., the instruction
pointer skips over them, or else {\em neutral} implying that their
execution will typically not affect the behavior of the program.
Trivial neutral instructions often involve the {\tt nop} instructions
(see Table I) which perform no function on their own when executed,
but do act to modify other instructions.  Thus, even though their
execution is neutral their particular value can still severely affect
the functioning of the organism. A perfectly neutral position sports
any of the $\D$ instructions with equal probability among a population
of sequences, while a maximally fixed position can only have one of
the $\D$ instructions there. To distinguish these, we define the

{\bf Per-Site Entropy} of a locus by trying out each of the $\D$
instructions at that position and evaluating the fitness of the
resulting organisms. All neutral positions are assigned an equal
probability to be expected at that site, while deleterious mutations
are assigned a vanishing probability (as they would be selected
against). Due to the uniform assignment of probabilities, the per-site
entropy of locus $x_i$ (normalized to the maximum entropy $\log(\D)$)
is
\begin{equation}
  H(x_i) = \frac{\log N_{\rm neut}(x_i)}{\log(\D)}\;.
\end{equation}
In an equilibrated population, this theoretical value of the per-site
entropy is a good indicator for the actual per-site entropy, measured
across the population (if the population is large enough).  As
positive mutations are extremely rare and we are only interested in
the diversity of the population when it is in equilibrium, for the
purposes of this
measurement they are treated as if they were neutral.  An indicator
for the randomness within a sequence is the

{\bf Per-Genome entropy}, which we approximate by the sum of the
per-site entropies
\begin{equation}
H = \sum_i^\ell H(x_i)\;.
\end{equation}
The actual per-genome entropy is in fact smaller, as the above
expression neglects {\em epistatic} effects which lead to correlations
between sites. For most purposes, however, the sum of the per-site
entropies is a good approximation for the randomness. Measuring the
entropy of the population by recording the individual genomic
abundances is fruitless as the sampling error is of the order of the
entropy~\cite{BASH59}. 
    
\section{Single Expression vs. Multiple Expression}
\begin{figure}[tb]
\centerline{\psfig{figure=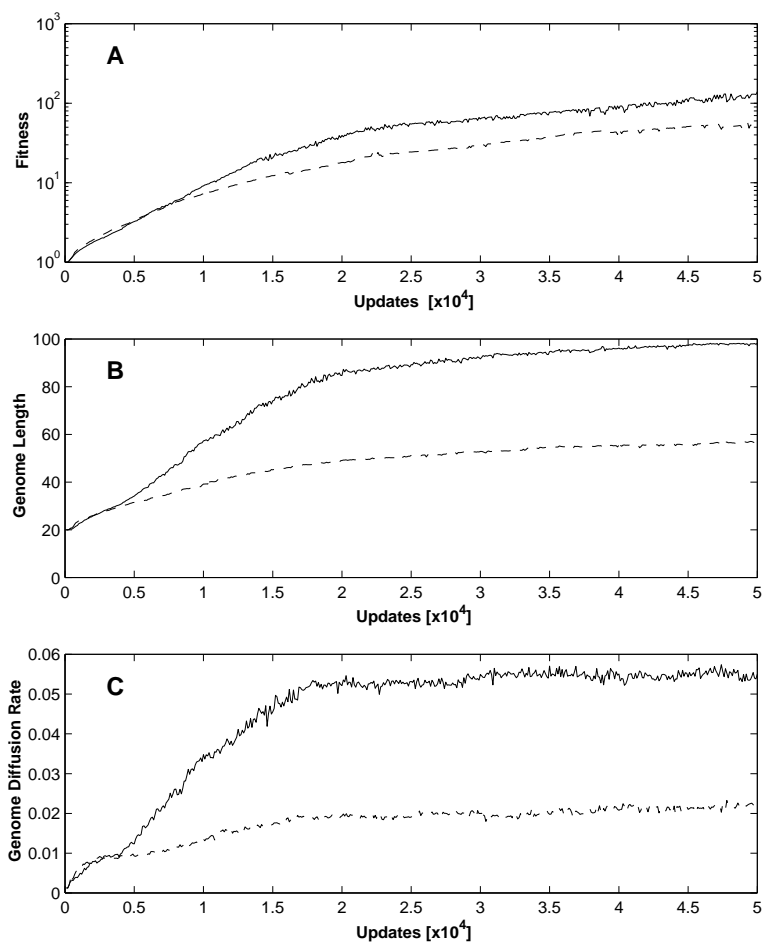,width=4.0in,angle=0}}
\caption{(A): Average fitness as a function of time (in updates) for 200
  populations evolved from $\ell=20$ ancestors, their average sequence
  length (B) and the average genomic diffusion rate (C) for the single
  expression chemistry controls (solid line) and the multiple expression
  chemistry (dashed line).}
\label{firstfig}
\end{figure}

Let us first examine adaptability as measured by the average increase
in fitness for both single and multiple expression chemistries. In
Fig.~\ref{firstfig}A, the fitness is averaged for the 200 trials\footnote{Each
  trial is seeded with a single ancestor, which quickly multiplies to
  reach the maximum number of programs in the population, set to 3,600
  for these trials. The population was subjected to copy mutations at
  a rate of $7.5\times 10^{-3}$ per instruction copied, and a rate of
  0.5\% of single insert or delete mutations per gestation period.}
which were seeded with small ($\ell=20$) seed sequences and no size
constraint (set I), for each of the chemistries.  
While the average increases relatively smoothly in time\footnote{Time
  is measured in arbitrary units called {\em updates}. Every update
  represents the execution of an average of 30 instructions per
  program in the population.}, it should be noted that each
individual fitness history is marked by periods of stasis interrupted
by sharp jumps, giving rise to a ``staircase'' picture reminiscent of
the adaptation of {\it E. coli}~\cite{ELE96}.  During adaptation, the
sequence length increases commensurately with the acquired
information, as shown in Fig.~\ref{firstfig}B. 
 
Clearly, the trials in which multiple expression is possible adapt
more {\em slowly} than the single-expression controls, a behavior that
may appear at first glance to be paradoxical as the only difference in
the underlying coding of the multiple expression trials is an {\em
  increased} functionality.  However, as we have noted previously, the
neutral fidelity of an organism directly determines the fraction of
its offspring which are viable.  As this value is inversely
correlated to the length of the genome, there is a pressure for the
genomes to evolve towards shorter length.  Normally, this pressure is
counteracted by the adaptive forces which require the organism to
store more information in its genome, requiring increased length.
Overlapping expression patterns (here, multiple parallelized
execution) allows this adaptation to occur while minimizing the length
requirement. Hence, multiple-expression genomes adapt more slowly.

The pitfalls of compacting so much information into the same portion
of the genome are illustrated in Fig.~\ref{firstfig}C where we plot the average
genomic diffusion rate $D_g$ for both chemistries.  It is evident in
this graph that initially both sets of experiments explore genetic
space at a comparable rate, but around approximately 5000 updates (on
average) the diffusion rates diverge markedly, followed by a
corresponding divergence in the fitness of the organisms (that a
higher diffusion rate leads directly to higher fitness in an
information-rich environment is shown in~\cite{ACO98}.) Investigating
the course of evolution further, we see that it is precisely at this
point that the differentiated, yet overlapping, use of multiple threads
is typically established.

To further implicate overlapping expression in reduced adaptation for
the populations, we consider (as was done in Ref.~\cite{MIYA78} for
the bacteriophage $\Phi X174$)) the substitution rate of instructions
for overlapping versus non-overlapping genes. The substitution rate in
\avida is equal to the neutrality (at equilibrium).  We find the {\em
  substitution suppression} (the neutrality in multiply expressed code
divided by the neutrality in singly expressed code) to be between 0.53
and 0.56 for the three sets of trials (Table II),
similar (but not quite as severe) as the suppression ratio of between
0.4 and 0.5 observed in the bacteriophages~\cite{MIYA78}. This was to
be expected, as there are no reading frames in \avida which implies
that two non-differentiated threads do not constrain the evolution any
more than a single thread. When the instruction pointers do adapt
independently and the threads differentiate, neutrality is
compromised.  Consequently, the instructions within sections of
overlapping code are comparatively ``frozen'' into their state.

\begin{table}[h]
\centering
\caption{Average neutrality of the final dominant genotype:  
  multiply-expressed code (column 1), singly expressed code (column
  2), and their ratio (column 3), for
  200 populations grown from $\ell=20$ ancestors (variable length)
  [set I], 100 populations grown from $\ell=80$ ancestors (variable
  length) [set II], and 100 populations grown from $\ell=80$ ancestors
  (constant length) [set III].}  \vskip 0.5cm
\begin{tabular}{|c||c|c|c|}
\hline Set & $\nu_{\rm mult}$ & $\nu_{\rm single}$ & ratio \\ \hline
I   & 0.109  & 0.202  & 0.539 \\
II  & 0.197  & 0.346  & 0.569 \\
III & 0.082  & 0.145  & 0.566 \\ \hline
\end{tabular}
\end{table}

\section{Evolution of Differentiation}
Let us now track the evolution of differentiation in more detail. 
We first address the {\em de novo} evolution of multiple expression,
i.e., the development of multi-threading from linear execution. 
This question has previously been addressed within {\sf tierra}~\cite{RAY91},
a population of self-replicating computer programs that served as the
inspiration to our {\sf avida}. In initial experiments, usage of
multiple threads would not evolve spontaneously, but hand-written programs
that had secondary expressions would evolve towards multiple
expression~\cite{RAY94}. More recently, experiments were carried
out within a network version of the {\sf tierra} architecture, which
showed that a program which used different instruction pointers to
execute different genes would not lose this ability~\cite{RAY98}. The
failure of multiple expression to evolve spontaneously in this system
can be tracked back to problems with {\sf tierra}'s digital chemistry
and the lack of an information-rich environment~\cite{OHA98}.

\begin{figure}[tb]
\centerline{\psfig{figure=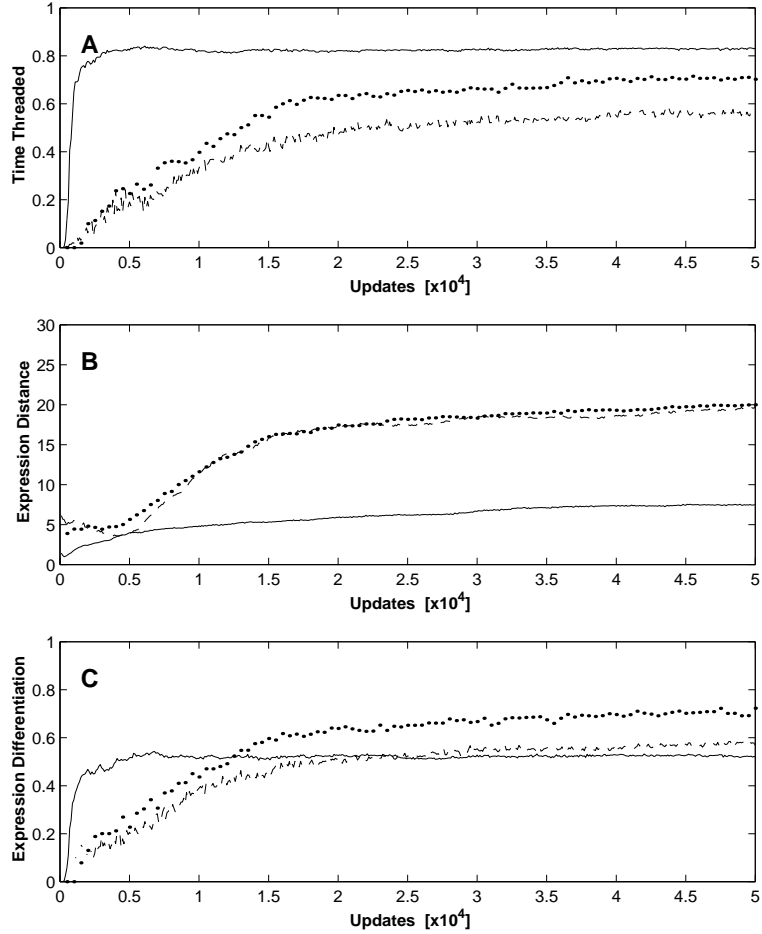,width=4in,angle=0}}
\caption{Differentiation measures. (A): Average fraction of lifetime 
  spent with secondary expression, as a function of time (in
  updates), (B): average expression distance, (C): average expression 
  differentiation. Set I (solid line), set II (dashed line), and set
  III (dotted line).}
\label{secondfig}
\end{figure}

Within {\sf avida}, the ability to use more than a single thread
begins to develop within the first 5000 updates and is very common
after about 10,000 updates, depending on the experimental boundary
conditions. Fig.~\ref{secondfig}A shows the (averaged) percentage of a
program's lifetime in which more than one thread is active, for the
populations of set I (solid line), set II (dashed line) and set III
(dotted line). It is apparent that multiple expression develops much
more readily in smaller genomes, due to the fact that the logistics are
less daunting.

\begin{figure}[tb]
\centerline{\psfig{figure=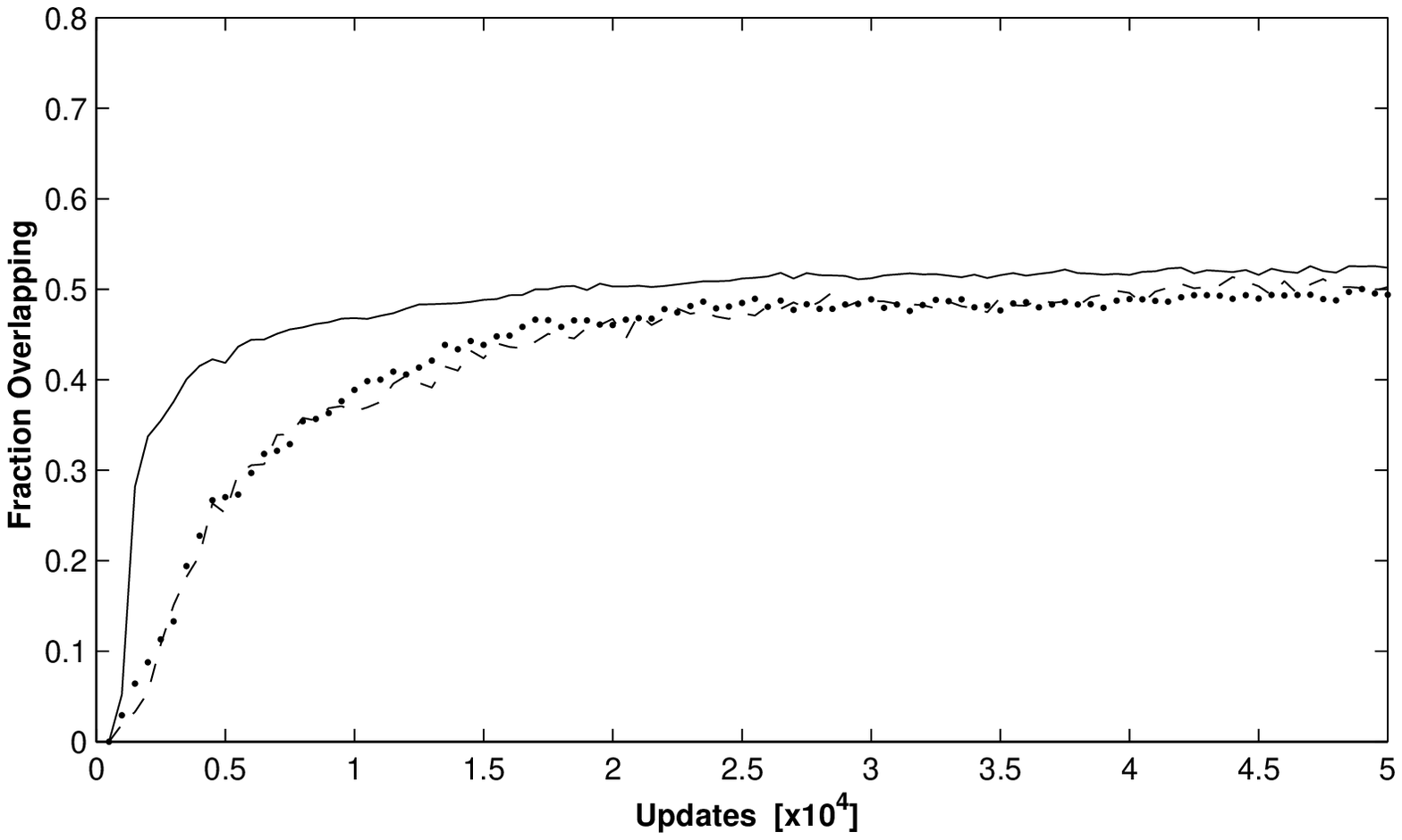,width=4in,angle=0}}
\caption{Average fraction of doubly expressed code for the three
  experimental sets. Solid line: set I, dashed line: set II, dotted
  line: set III.}
\label{thirdfig}
\end{figure}

In panels B and C of Figure \ref{secondfig} we display two indicators of
differentiation (defined earlier), the expression distance and the
expression differentiation, respectively. The expression distance
appears to be sensitive to the experimental starting condition, as set II and
set III show a value over twice that of set I. We observe that this is
due to the small size of the ancestor used in set I: as that ancestor
develops threading very quickly, it loses adaptability earlier and
lags both in average fitness and average sequence length. In fact,
those averages are dragged down by a significant percentage of the
trials in set I which were stuck in an evolutionary dead-end. Set II
and III were seeded with an ancestor of length $\ell=80$ and did not
suffer from this lot. Fig.~\ref{secondfig}C shows the {\em expression
  differentiation}, i.e., the fraction of code that is executed
differently by the two threads. This fraction is less dependent on
experimental conditions, and the genomes appear to develop towards 0.5. 
Note, however, that this measure cannot accurately reflect
differentiation which is more subtle than threads executing particular
instructions a different number of times. For example, two threads
which execute a stretch of code in an identical manner but that start
execution at different points ``upstream'' may end up calculating
very different functions, and thus have quite different behaviors. This
difference will thus be underestimated. While the preceding graphs
seem to indicate that differentiation stops about half-way through the
duration we record, this is actually not so, as the more microscopic analysis 
of the following section reveals. Finally, Fig.~\ref{thirdfig} shows
the evolution of the fraction of code that is executed by multiple
threads. 
 
We anticipate that this fraction rises swiftly at first, but
then levels off, as it is not advantageous to multiply express all
genes (see below). However, we might anticipate that the fraction
would start to decline at some point, when the organism develops the
ability to localize its genes and use independent instruction pointers
for each of them. We do not witness this trend in Fig.~\ref{thirdfig}
presumably because there is no cost associated with the development
of secondary expression. This should be viewed as a peculiarity of the
digital environment rather than a universal feature, which we hope to
eliminate with future refinements of the \avida world.

\section{Evolution of Genetic Locality}

To get a better idea of how evolution is acting upon programs
harboring multiple threads, we must look at exactly what is being
expressed.  We can loosely characterize all organisms by tracking
three separate genes.  They are ``self-analysis'' ({\it slf}),
``replication'' ({\it rpl}) , and ''computation'' ({\em cmp}).  To
follow the progression of these genes through time, we examine a
sample experiment seeded with an ancestor of size 80 (as before,
capable only of self-replication), in an environment in which
size-altering mutations are strictly forbidden (a trial from set III).
This limitation was enforced in order to better study the
functionality of the organism and the location of its genes.  Similar
studies have been done with all 400 trials used to collect the bulk of
the data for this report, showing comparable behavior.

\begin{figure}[tb]
\centerline{\psfig{figure=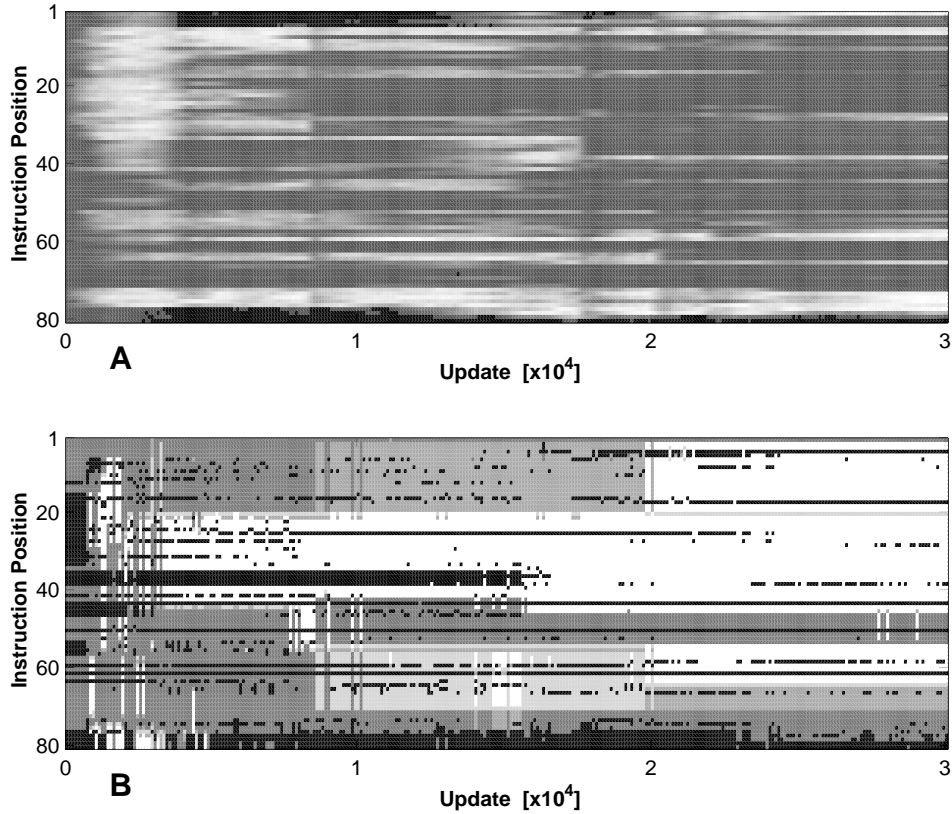,width=5in,angle=0}}
\caption{(A): Per-site entropy for each locus as a function of time
  for a standard (set III) trial. Random (variable) positions with near-unit
  per-site entropy are bright, while ``fixed'' instruction with
  per-site entropy near zero are dark. (B): Thread identification
  within a genome. Black indicates instructions which are never directly
  executed, dark grey denotes instructions executed by a single thread
  when no other thread is active, while sections which are executed by
  a single thread while another thread is executing a different
  section are colored in lighter shades of grey. Sections with
  overlapping expression are in white.}
\label{fourthfig}
\end{figure}

In Fig.~\ref{fourthfig}A we follow the per-site entropies for each locus as a
function of time. Positions are labeled by 1 to 80 on the vertical
axis, while time proceeds horizontally. A grey-scale coding has been
employed to denote the variability of each locus, where the white end
denotes more variable positions and the dark end more fixed
positions. Because the per-site entropies have been calculated by
obtaining the frequency with which each instruction appears at that
locus within the population (as opposed to the theoretical estimate
based on neutrality), major evolutionary transitions are identifiable
by dark vertical bands. 
Fig.~\ref{fourthfig}B shows which portion of the code is expressed by
which pointer, by two pointers simultaneously, or not at all. 

The first gene {\it slf} uses pattern matching on {\tt nop}
instructions in order to find the limits of its genome and from that
calculate it's length.  This value is used for {\it elongation} (via
the command {\tt alloc}), which adds empty memory to the genome and
prepares it for the ``execution'' of the replication gene. Note that
avidian genomes are circular. There are
two interesting points to note about the evolution of {\it slf}:
First, there are many methods by which the organism can determine its
own genomic length, so this gene tends to vary widely.  Most of the
time the organism keeps pattern matching techniques, but matches
different portions of the code. However, often an organism shifts to
purely numerical methods performing mathematical operations upon
itself which yield the genome length ``by accident''.  The other
evolutionary characteristic of this gene is that there is no benefit
in expressing it multiple times as it has a fixed result which needs
only be applied once during the gestation cycle.  Looking at
Figure~\ref{fourthfig}, the {\it slf} gene initially spans from lines
44 to 61 plus the first four lines and last four lines of the genome
which are boundary markers fashioned from {\tt nop} instructions.  The
first major modification to the {\it slf} gene occurs around update
3000.  The pattern used to mark the limits of the genome is a series
of four {\tt nop-A} instructions.  As a newly allocated genome has all
of its sites initialized to {\tt nop-A}, the genome is re-organized
such that these lines are no longer copied.  This reduces the
possibility of variation in these sections of code to zero.
This is apparent in Fig.~\ref{fourthfig}A
as the positions of these limit patterns become completely black
indicating vanishing entropy.

The {\it slf} gene is continuously undergoing minor changes as is
becomes more optimized to require fewer lines of code to perform its
function.  Near update 13,000 it shifts dramatically and is replaced
by one in which size is calculated using only the final boundary
markers.  The distance from the gene to the final marker is
determined, and then manipulated numerically in order obtain the
number which is the size of the organism.  Looking at the first four
lines of Fig.~\ref{fourthfig}A around this update, we see that they
are slowly phased out and increase in entropy as they are no longer as
critical to the organism's survival.  Finally the size of the pattern
marking the end boundary of the organism is shortened until it becomes
only a single line.  By the end of the evolution shown, the {\it slf}
gene only occupies lines 48 through 56. Note that all of these lines
are only expressed a single time.

The next gene under consideration is the actual replication gene {\it
  rpl}.  This sequence of instructions uses the blank memory allocated
in the self-analysis phase and enters a 'copy-loop' which moves line
by line through the genome, copying each instruction into the newly
available space.  When this process is finished, it severs the newly
created copy of itself which is then placed in an adjacent lattice
site.  These dynamics spawn off a new organism which, if the copy
process was free of mutations, would be identical to the parent.  In
Fig.~\ref{fourthfig}, the organism being tracked has its replication
gene on lines 65 to 71 until update 24,000 at which time this gene
actually grows an additional line becoming much more efficient by
``unrolling'' its copy-loop.  What this means is that it is now able
to copy {\em two} lines each time through the loop.  From the dark
color of these lines, it is obvious that they have very low entropy,
and are therefore very difficult to mutate.  The copy-loop is a very
fragile portion of code, critical to the self-replication of the
organism, yet we do see some evolution occurring here when multiple
threads are in use.  Often the secondary thread will simply ``fall
through'' the copy-loop (not actually looping through to copy the
genome) and move on to the next gene, while the other thread performs
the replication.  However, sometimes the two threads will actually
evolve together to use the copy loop in different ways, with each
thread copying {\em part} of the genome. In Fig.~\ref{fourthfig}, most
of the {\em rpl} gene is executed by only one thread. The {\em rpl}
gene is followed by junk code which, while executed sporadically, does
not affect the fitness in any way (as evidenced by the light shading
in Fig.~\ref{fourthfig}A for these lines).

The most interesting of the genes is the computation gene {\it cmp}.
The ancestor does not possess this gene at all, so it evolves
spontaneously during the adaptive process.  There are 78 different
computations rewarded in this environment, all of which are based on
bit-wise logical operations.  The organisms have three main commands
which they use to accomplish those: a {\tt get} instruction which
retrieves numbers from the environment, a {\tt put} instruction to
return the processed result, and a {\tt nand} instruction which
computes the logical operation not-and (see Table I).  Any logical
operation can be computed with a properly arranged
collection of {\tt nand} instructions.

The {\it cmp} gene(s) evolve uniquely in each trial, enabling the
organisms to perform differing sets of tasks.  There are, however,
certain themes which we see used repeatedly whereby the same section
of code is used by both threads, but their initial values (i.e., the
processing performed thus far on the inputs) differs. Consequently,
this section of code performs radically different tasks, actually
encouraging this overlapping.  Portions of this algorithm which might
have some neutrality for a single thread of execution will now be
frozen due to the added constraints imposed by a secondary execution.
The size of {\it cmp} grows during adaptation as a number of
computations are performed, and the gene is almost always expressed by
both threads as this is always advantageous. In Fig.~\ref{fourthfig},
the {\it cmp} gene stretches from line 1 to line 42 (at update
30,000), while it is considerably smaller earlier. Furthermore, the
genome manages to execute the entire gene by both threads (the
transition from single expression of part of {\it cmp} to double
expression is visible around update 20,000).  This gene ends up being
expressed many times (as the instruction pointers return to this
section many times during execution). All in all, 17 different logical
operations are being performed by this gene.

By the end of the evolution tracked in Fig.~\ref{fourthfig}, most of
the genes appear to occupy localized positions on the genome. The {\it
  cmp} gene (white sections in Fig.~\ref{fourthfig}) is revisited many
times by both threads with differing initial conditions for the
registers, allowing the genome to maximize the computational
output. In the meantime, those sections have become fixed (their
variability is strongly reduced) as witnessed by their dark shading in
Fig.~\ref{fourthfig}A. 

\section{Discussion and Conclusions}

The path taken by evolution from simple organisms with few genes
towards the expression of multiple genes via overlapping and
interacting gene products in complex organisms is difficult to retrace
in biochemistry.  Artificial Life, the creation of living systems
based on a different chemistry but using the same universal principles
at work in biochemical life, may help to understand some key
principles in the development of gene regulation and the organization
of the genetic code. We have examined the emergence and
differentiation of code expression {\it in parallel} within a digital
chemistry, and found some of the same constraints affecting multiply
expressed code as those observed in the overlapping genes of simple
biochemical organisms. For example, multiply expressed code is more
fragile with respect to mutations than code that is ``transcribed'' by
only one instruction pointer, and as a result evolves more slowly.
During most stages of evolution, two constraints are most notable: the
pressure to reduce sequence length in order to lessen the mutational load, and
the pressure to {\em increase} sequence length in order to be able to
store more information. Simple organisms can give in to both pressures
by using overlapping genes, gaining in the short term but mortgaging
the future: the reduced evolvability condemns such organisms to a
slower pace of adaptation, and exposes them to the risk of extinction
in periods of changing environmental conditions.

This trend is clearly visible in the evolution of digital organisms,
as is a trend towards multiple expression of as much of the code as
possible. This latter feature we believe {\em not} to be universal,
but rather due to the fact that multiple expression in \avida is
cheap, i.e., no resources are being used in order to express more
code. In a more realistic chemistry, this would not be the case:
adding an instruction pointer should put some strain on the organism
and use up energy; in such circumstances multiple expression would
only emerge if the advantage of the secondary expression outweighs the
cost of it. We also expect more complex gene regulation in such an
environment, as genes would be turned on only when needed. 

Still, under extreme conditions we believe that multiple overlapping
genes are a standard path that any chemistry might follow. Even though
evolution slows down, such organisms can be rescued either by the
development of error-correction algorithms, or an external change in
the error rate. In either case, a drastic reduction of the mutational
load would enable the sequence length to grow and the overlapping
genes to be ``laid out'' (for example by gene-duplication). The
corresponding easing of the coding constraints might give rise to an
explosion of diversity and possibly the emergence of
multi-cellularity.\\

\noindent
{\bf Acknowledgements.}  We would like to thank Grace Hsu and Travis
Collier for collaboration in the initial stages of this work.  Access
to a Beowulf system was provided by the Center for Advanced
Computing Research at the California Institute of Technology.  This
work was supported by the National Science Foundation.

\bibliographystyle{amsalpha}

\end{document}